\documentclass[9pt]{article}

\usepackage[a4paper, bindingoffset=0cm, hmargin={2cm, 2cm},vmargin={2cm, 4cm}]{geometry}

\usepackage[utf8]{inputenc}
\usepackage[T1]{fontenc}
\usepackage{lmodern}
\usepackage[english]{babel}
\usepackage{amsmath}
\usepackage{amsfonts}
\usepackage{authblk}
\usepackage{color}
\usepackage{amssymb}
\usepackage{graphicx}
\usepackage{fancyhdr}
\usepackage[notrig]{physics}
\usepackage{multicol}

\usepackage[sorting=none,style=nature, backend=biber]{biblatex}
\usepackage{titlesec}
\titleformat*{\section}{\normalsize\bfseries}

\usepackage[font=small,labelfont=bf]{caption}

\usepackage{indentfirst}
\usepackage[colorlinks=true, linkcolor=black, citecolor=black, urlcolor=black]{hyperref}

\usepackage{float}
\restylefloat{figure} 

\usepackage{subcaption}

\usepackage{tikz,xcolor,hyperref}
\definecolor{lime}{HTML}{A6CE39}
\DeclareRobustCommand{\orcidicon}{
	\begin{tikzpicture}
	\draw[lime, fill=lime] (0,0) 
	circle [radius=0.16] 
	node[white] {{\fontfamily{qag}\selectfont \tiny ID}};
	\draw[white, fill=white] (-0.0625,0.095) 
	circle [radius=0.007];
	\end{tikzpicture}
	\hspace{-2mm}
}
\foreach \x in {A, ..., Z}{
	\expandafter\xdef\csname orcid\x\endcsname{\noexpand\href{https://orcid.org/\csname orcidauthor\x\endcsname}{\noexpand\orcidicon}}
}

\title{Radiation Emission during the Erasure of Magnetic Monopoles}
\author{Maximilian Bachmaier\orcidB{}\footnote{maximilian.bachmaier@physik.uni-muenchen.de}}
\author{Gia Dvali}
\author{Juan Sebasti\'an Valbuena-Berm\'udez\orcidA{}\footnote{juanv@mpp.mpg.de}}
 \affil{\small \textit{
 	Arnold Sommerfeld Center,
 	Ludwig-Maximilians-Universit{\"a}t,
 	Theresienstra{\ss}e 37,
 	80333 M{\"u}nchen,
 	Germany}}
 \affil{\small \textit{
 	Max-Planck-Institut f{\"u}r Physik,
 	F{\"o}hringer Ring 6,
 	80805 M{\"u}nchen,
	Germany}}
\date{\small\today} 

\begin{document}
    \normalsize
    \maketitle

    \begin{abstract}
    \noindent
        We study the interactions between 't Hooft-Polyakov magnetic monopoles and
        the domain walls formed by the same order parameter within an $SU(2)$ gauge theory. 
        We observe that the collision leads to the erasure of the
        magnetic monopoles, as suggested by Dvali, Liu, and Vachaspati \cite{Dvali-Liu-Vachaspati:1997}. The domain wall represents a layer of vacuum with un-Higgsed $SU(2)$
        gauge symmetry. 
        When the monopole enters the wall, it unwinds, and the magnetic charge spreads over the wall. We perform numerical simulations of the collision process and in particular analyze the angular distribution of the emitted electromagnetic radiation. As in the previous studies, we observe that erasure always occurs. Although not forbidden by any conservation laws, the monopole never passes through the wall. This is explained by entropy suppression. The erasure phenomenon has important implications for cosmology, as it sheds a very different light on the monopole abundance in post-inflationary phase transitions and provides potentially observable imprints in the form of electromagnetic and gravitational radiation. The phenomenon also sheds light on fundamental aspects of gauge theories with coexisting phases, such as confining and Higgs phases.
    \end{abstract}

    \begin{multicols}{2}
        \section{Introduction}
\label{sec:introduction}
Topological defects play significant roles in different branches of physics. These entities emerge in theories with topologically non-trivial vacuum manifolds. In particular, such manifolds are common in theories with spontaneously broken symmetries. When symmetry breaking takes place in a phase transition during the cosmological evolution, the defects can be formed via the Kibble mechanism~\cite{Kibble:1976sj}.

In~\cite{Dvali-Liu-Vachaspati:1997}, it has been pointed out 
that defects can be subjected to a so-called 'erasure' phenomenon. Namely, in some cases, one and the same order parameter simultaneously gives rise to defects of different dimensionality, e.g., magnetic monopoles and domain walls.

In such cases, upon the encounter, the less extended defects can be erased by the more extended ones.  
In~\cite{Dvali-Liu-Vachaspati:1997} this effect was discussed
for the domain walls and magnetic monopoles system. 
In particular, it was pointed out that the grand unified phase transition, which ordinarily creates 
't Hooft-Polyakov 
magnetic monopoles~\cite{tHooft:1974, Polyakov:1974}, 
can also give rise to domain walls. Upon the encounter, 
the magnetic monopole is erased by the domain wall. 
The essence of the erasure is that the domain wall 
creates a supporting surface for unwinding the monopole 
field. The Higgs field vanishes inside the wall. 
Due to this, the magnetic charge, instead of 
staying localized at a point, spreads over the entire wall.

Original motivation of~\cite{Dvali-Liu-Vachaspati:1997}
was the solution to the cosmological magnetic
monopole problem~\cite{Preskill:1979, Zeldovich:1978}.
The idea is that the domain walls 'sweep away' monopoles and disappear.  
For short, we shall refer to this dynamics as the 
DLV mechanism. It was already a subject of numerical studies in~\cite{Pogosian-Vachaspati:2000, Brush-Pogosian-Vachaspati:2015}.

The monopole erasure scenario allows 
to have the monopole production after the 
inflationary phase without conflicting with the constraints 
on the monopole abundance. 
It therefore 'liberates' the grand unified symmetry 
from the necessity of being broken during inflation. 
This is beneficial for some
motivated inflationary scenarios predicting the
grand unified phase transition after inflation. 

The DLV erasure mechanism plays an important role in 
generic quantum field theoretical systems with defects supporting different gauge theories' phases.
An early example is provided by a confining 
gauge theory (e.g.~$SU(2)$)  which contains domain walls with de-confined $U(1)$ Coulomb phase of the same gauge interaction~\cite{Dvali-Shifman:1996}. Due to confinement, in the 
$SU(2)$ vacuum, the gauge electric field is trapped in 
the form of QCD flux tubes. 
However, the wall serves as a base for the spread-out of the 
QCD electric flux. Correspondingly, for the QCD string, the wall plays a role similar to a $D$-brane. Upon encountering 
such a wall, the QCD string gets erased~\cite{Dvali-Vilenkin:2002, Dvali-Nielsen-Tetradis}. 
The dual version of this, in the form of the erasure of 
vortices and strings by a domain wall, was recently studied numerically in~\cite{Dvali-Valbuena:2022}.

One important general question is the efficiency of the erasure. 
As suggested in the work on the monopole-wall system~\cite{Dvali-Liu-Vachaspati:1997}, the 
erasure mechanism was expected to be very efficient. 
Although topologically, it is allowed for a monopole 
to pass through the wall, this passage is expected to be 
highly improbable.
The argument of DLV was based on loss of coherence 
in the monopole wall collision. Namely, upon collision
with the wall, the monopole charge starts to spread in the traveling waves. This makes the further recombination of the monopole on the other side of the wall very unlikely. 
As supporting evidence for this reasoning, in~\cite{Dvali-Liu-Vachaspati:1997} the  
results of numerical studies of interactions between the 
skyrmions and walls~\cite{Kudryavtsev:1997nw, Kudryavtsev:1999zm} were used.

In more recent studies, the efficiency 
of the erasure phenomenon was repeatedly observed 
in monopole-anti-monopole~\cite{Dvali-Valbuena-Zantedeschi:2022}, 
wall-vortex, and string-wall~\cite{Dvali-Valbuena:2022} systems. The analytic explanation of these numerical results
was given by substantiation of the DLV coherence loss argument \cite{Dvali-Liu-Vachaspati:1997} by the entropy-count of~\cite{Dvali:2020}.
This count indicates that the probability  
of survival is exponentially suppressed due to the fact that 
the final state after erasure has a much higher entropy 
in comparison to a surviving defect.

In the present paper, we extend the study of the erasure 
phenomenon in the
monopole-wall system. We use a simple prototype 
model with an adjoint Higgs field of $SU(2)$  
which possesses $U(1)$ invariant vacua separated by domain walls 
(vacuum layers) with $SU(2)$ invariant phases. 
The monopoles that exist in the $U(1)$ phase
get erased upon the encounter with the domain walls 
that support the $SU(2)$ phase in their interior. 
Again, we observe that the erasure occurs for the considered parameters regime.

The main novelty is the analysis of the emitted electromagnetic radiation during the erasure. The emission of electromagnetic radiation accompanies the spread-out of the magnetic charge of the monopole. 
This can have several interesting implications both for the theoretical understanding of the erasure phenomenon as well as for its observational consequences.

        \section{Generalities of Radiation}
\label{sec:generalities-of-radiation}

Let us review a fundamental phenomenon of classical electrodynamics that will become relevant to our discussion.
It is a well-known fact that the acceleration of electric charge leads to an emission of radiation.
If we allow the existence of magnetic charges, the acceleration of magnetic charges will lead to the same effect due
to the duality of the extended Maxwell equations.
The behavior of the electric and magnetic fields, albeit, is exchanged.
Consider a point charge $q$ located at the origin with initial velocity $u$ and acceleration $a$, where the direction of the velocity is parallel to the direction of acceleration.
The energy density of the radiation for this situation can be calculated analytically and is given by~\cite{Griffiths:2017}
\begin{align}
\label{eq:energy-radiation-pattern}
  \varepsilon = \frac{q^2}{16\pi^2}\frac{a^2}{r^2}\frac{\sin^2 \theta}{\left(1-u\cos\theta\right)^6},
\end{align}
where $r$ is the distance from the charge and $\theta$ is the angle relative to the direction of movement.
As we can deduce from this equation, the energy density is not distributed homogeneously on a sphere around the point charge.
Most of the radiation gets emitted in the direction
\begin{align}
    \label{eq:angle_maximal_radiation}
  \theta_{\max}=\arccos\left(\frac{-1+\sqrt{1+24 u^2}}{4 u}\right).
\end{align}

Furthermore, we can notice that the form of the distribution depends only on the initial velocity of the charge and does not depend on the acceleration.
The shape of the radiation emission is depicted in figure~\ref{fig:radiation_pattern_theory} using a normalized radiation pattern.
The greater the initial velocity, the more the loops bend in the direction of the initial motion.

We observe that the direction of radiation emitted during the erasure of a magnetic monopole is comparable to the expected one for a constant accelerated magnetic point charge. We will elaborate on this point below.\\
\begin{minipage}{\linewidth}
  \includegraphics[width=\linewidth]{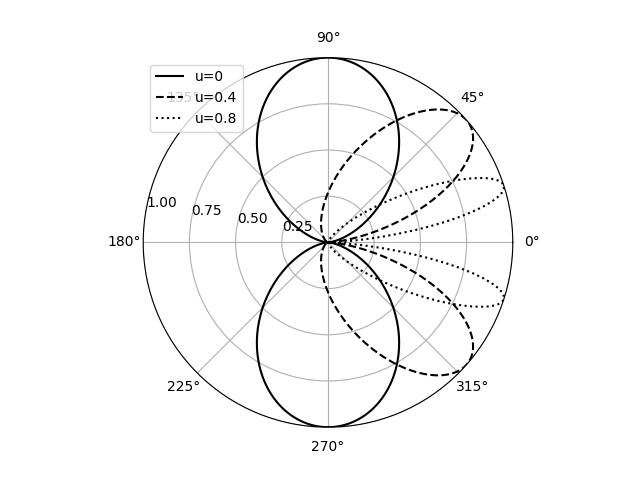}
  \captionof{figure}{The radiation pattern for an accelerated charge with initial velocity $u$. The radius represents the normalized value of the radiation energy density   $\frac{\varepsilon}{\varepsilon(\theta_{\max})}$.}
  \label{fig:radiation_pattern_theory}
\end{minipage}
        \section{The Model and its Solutions}\label{sec:the-model-and-its-solutions}
We consider a model with an $SU(2)$ gauge symmetry and a scalar field $\phi$, transforming under the adjoint representation. 
This model is a prototype of grand unified theories, which is still able to capture the essence of the occurring phenomena.
The Lagrangian is given by~\cite{Dvali-Nielsen-Tetradis}
\begin{align}
    \label{eq:Lagrangian}
  \mathcal{L}=-\frac{1}{2}\Tr\left(G_{\mu\nu}G^{\mu\nu} \right)+\Tr\left((D_\mu\phi)^\dagger (D^\mu\phi) \right)-V(\phi),
\end{align}
with the potential
\begin{align}
  \label{eq:potential}
  V(\phi)=\lambda\left(\Tr(\phi^\dagger\phi)-\frac{v^2}{2}\right)^2 \Tr(\phi^\dagger\phi).
\end{align}
Notice that $\lambda$ has the mass dimension $-2$. The scalar field can be written as $\phi=\phi^a T_a$, where the $SU(2)$ generators $T_a$ are normalized
as $\Tr(T_a T_b)=\frac{1}{2}\delta_{ab}$.
The field strength tensor is defined by
\begin{align}
  G_{\mu\nu}\equiv\partial_{\mu}W_{\nu}-\partial_{\nu}W_{\mu}-ig \left[ W_\mu, W_\nu \right],
\end{align}
with the gauge fields $W_\mu\equiv W^a_\mu T_a$.
The covariant derivative has the usual form
\begin{align}
  D_{\mu}\phi\equiv\partial_{\mu}\phi-ig\left[ W_\mu,\phi \right].
\end{align}

The feature of the sextic potential is that it has two disconnected vacua, corresponding to the $SU(2)$ invariant
phase, $\langle \Tr(\phi^\dagger\phi)\rangle =0$,
and the phase with $SU(2)$ Higgsed down to $U(1)$, $\langle \Tr(\phi^\dagger\phi)\rangle = \frac{v^2}{2}$.

In the $SU(2)$ invariant vacuum, the vector fields are massless while $\phi$ is massive.
On the other hand, in the second vacuum, the symmetry group $SU(2)$ is Higgsed down to $U(1)$
and two of the vector fields gain
the mass $m_v=vg$ through the Higgs mechanism, while one stays massless. The mass of the Higgs boson is given by $m_h=\sqrt {\lambda}v^2$.

At the quantum level, the $SU(2)$ invariant vacuum becomes confining. However, for the considered parameters, this can be ignored. We will elaborate more on this later.
As a first approximation, let us consider the classical equations of motion.
They are given by
\begin{align}
  &\partial_{\mu}(D^\mu\phi)^a+g\varepsilon^{abc}\ W^b_{\mu} (D^\mu \phi)^c + \pdv{V}{\phi^a}=0,\label{eq:field-equation1}\\
  &\partial_{\mu}G^{a\mu\nu}+g\varepsilon^{abc}\ W^b_{\mu}G^{c\mu\nu}-g\varepsilon^{abc}\ (D^\nu \phi)^b \phi^c=0.\label{eq:field-equation2}
\end{align}

The spectrum of the model contains magnetic monopoles which are realized as solitons in the $U(1)$ vacuum.
Consider the 't Hooft-Polyakov ansatz~\cite{tHooft:1974,Polyakov:1974}
\begin{align}
  W^a_i&=\varepsilon_{aij}\frac{r^j}{r^2}\frac{1}{g}(1-K(r)),\nonumber\\
  W^a_t&=0,\nonumber\\
  \phi^a&=\frac{r^a}{r^2}\frac{1}{g}H(r),
\end{align}
thus, the field equations~\eqref{eq:field-equation1} and~\eqref{eq:field-equation2} reduce to
\begin{align}
  K''=&\frac{1}{r^2}\left(K^3-K+H^2 K\right),\nonumber\\
  H''=&\frac{2}{r^2}H K^2\nonumber\\
  &+m_h^2 \left(\frac{3}{4}\frac{1}{r^4 m_v^4}H^5-\frac{1}{r^2 m_v^2}H^3+\frac{1}{4}H\right).
\end{align}

To ensure good behavior at the boundary, the following standard boundary conditions are required
\begin{align*}
  &K(r)\xrightarrow{r\rightarrow 0}1 ,&& K(r)\xrightarrow{r\rightarrow\infty}0,\\
  &K'(r)\xrightarrow{r\rightarrow 0}0, && \frac{H(r)}{m_v r}\xrightarrow{r\rightarrow\infty}1,\\
  &\frac{H(r)}{m_v r}\xrightarrow{r\rightarrow 0}0.
\end{align*}
The profile functions $H(r)$ and $K(r)$ were found numerically by using an iterative method that starts
at the solution in the BPS limit $m_h \rightarrow 0$~\cite{Bogomolny:1975de, Prasad-Sommerfield:1975} and relaxes to the solution
with $m_h \neq 0$.
For the later simulations, we evaluated in this way the profile function
for $\frac{m_h}{m_v}=1$ (see figure~\ref{fig:profile_functions}).\\
\begin{minipage}{\linewidth}
    \includegraphics[width=\linewidth]{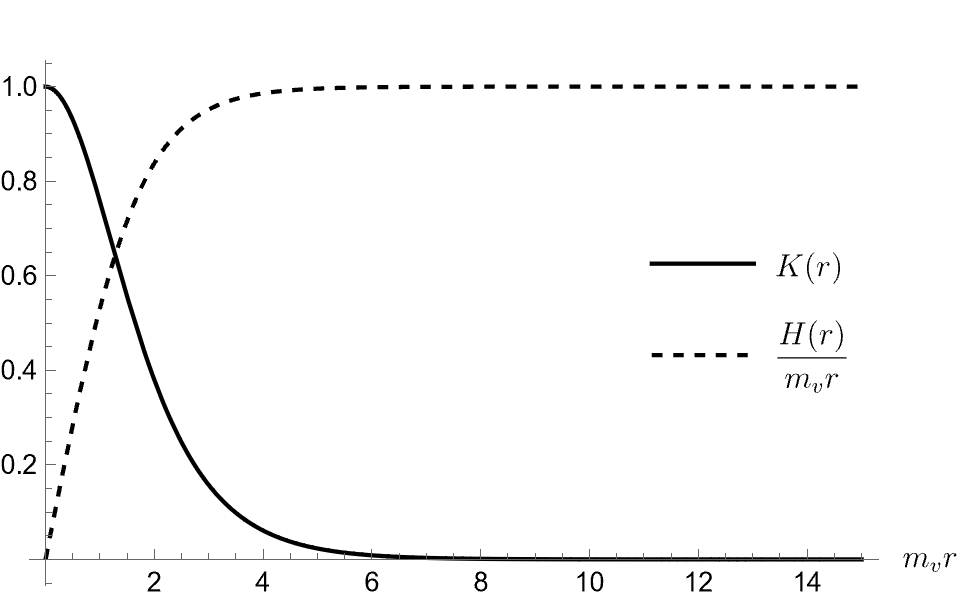}
    \captionof{figure}{The profile functions of the magnetic monopole for $\frac{m_h}{m_v}=1$.}
    \label{fig:profile_functions}
\end{minipage}\\[0.5cm]

As mentioned before, the present work aims to study the interaction between domain walls and magnetic monopoles.
We anticipate that the monopole is erased during the collision, and electromagnetic radiation is emitted in this process.
In order to analyze the radiation, we need to know the electric and magnetic fields.
Following the standard definitions, the non-abelian magnetic and electric fields can be written analogously to classical electrodynamics as
\begin{align}
  B^a_k&=-\frac{1}{2}\varepsilon_{kij}G^a_{ij},\\
  E^a_k&=G^a_{0k}.
\end{align}
Since we are interested in the $U(1)$ magnetic and electric fields, it is necessary to project out the component
that points in the direction of the electromagnetic charge operator $Q=\frac{\phi^a}{\sqrt {\phi^b \phi^b}}T^a$.
Using the scalar product $\langle A,B\rangle=2\Tr(AB)$ one can find
\begin{align}
  B^{U(1)}_k=\frac{\phi^a}{\sqrt {\phi^b \phi^b}}B^a_k,\label{eq:BU1}\\
  E^{U(1)}_k=\frac{\phi^a}{\sqrt {\phi^b \phi^b}}E^a_k.\label{eq:EU1}
\end{align}
The definitions \eqref{eq:BU1} and \eqref{eq:EU1} are valid in the $U(1)$ invariant phase, $\phi^b \phi^b=v^2$, and thus valid for long distances compared
to the size of the monopole core $\sim m_h^{-1}$.\\

The potential \eqref{eq:potential} allows the two phases to co-exist.
Therefore, we can treat the potential as an intermediate step of a first-order phase transition.
In the model \eqref{eq:Lagrangian}, domain walls interpolate between  $SU(2)$ and $U(1)$ invariant phases.
For a planar domain wall located at $z=0$ with ansatz $\phi^1=\phi^2=0$ and $\phi^3=\phi(z)$, the non-trivial solutions of the Bogomolny equation~\cite{Bogomolny:1975de}
\begin{align}
  \phi'=\pm \sqrt {2V},
\end{align}
derived from the field equation \eqref{eq:field-equation1} are 
\begin{align}
  \phi_{(\pm v,0)}(z)&=\frac{\pm v}{\sqrt{1+e^{m_h z}}},\\
  \phi_{(0,\pm v)}(z)&=\frac{\pm v}{\sqrt{1+e^{-m_h z}}}.
\end{align}
The boundary values of these solutions are on one side $\pm v$ and on the other side $0$, which correspond to the $U(1)$ invariant phase and the $SU(2)$ invariant phase, respectively.
In order to study the erasure mechanism, we consider the passage of a monopole through an $SU(2)$ invariant vacuum layer. The vacuum layer can be approximated as a combination of two parallel domain walls, for instance~\cite{Dvali-Valbuena:2022}
\begin{align}
  \phi_{\text{VL}}(z)=\phi_{(v,0)}(z)+\phi_{(0,v)}(z-h),
  \label{eq:VL-Ansatz}
\end{align}
where $h$ is the distance between the two domain walls.
Note that for finite $h$, the vacuum layer is not a solution to the static field equations, since the wall and anti-wall attract each other.
However, the interaction is negligible for $h\gg m_h^{-1}\sim m_v^{-1}$.
We used $h=20m_v^{-1}$ in the simulations.
In this regime, the vacuum layer is sufficiently long-lived during the period of investigation~\cite{Dvali-Valbuena:2022}.

        \section{Initial Configuration}\label{sec:initial-configuration}
We numerically study the interaction of a magnetic monopole and an $SU(2)$ invariant vacuum layer.
To achieve this, we numerically solved the equations~\eqref{eq:field-equation1} and~\eqref{eq:field-equation2}.
As initial configuration, we considered field configurations where the vacuum layer is Lorentz boosted towards the monopole.
Upon the collision, we bear out the 'sweeping away' mechanism~\cite{Dvali-Liu-Vachaspati:1997}.
In particular, we observed that the monopole is unable to pass the layer; instead, the magnetic charge dissolves and spreads out.
Additionally, electromagnetic radiation gets emitted.
As mentioned before, we expect the form of the radiation pattern to depend on the initial velocity of the magnetic charge.
This anticipation prompted us to elaborate on situations where the magnetic monopole is also Lorentz boosted.
Furthermore, boosting the magnetic monopole simultaneously with the vacuum layer allows us to
check the mechanism for much higher collision velocities.

The maximal velocities we could study with an appropriate accuracy were $0.8$ (in units of $c=1$) for the magnetic monopole
and $0.98$ for the vacuum layer.
For higher velocities, the resolution of the lattice was not acceptable.
These two cases allow us to check the erasure mechanism for Lorentz factors of $\gamma_\text{M}=1.67$ and $\gamma_\text{VL}=5.03$, respectively.
Boosting both objects with these velocities albeit leads to the collision relative speed of about $0.9977$, where we used the addition
rule for relativistic velocities $u=\frac{u_1+u_2}{1+u_1 u_2}$.
Therefore, we were able to check the erasure mechanism for the ultra-relativistic regime up to a gamma factor of about $\gamma=15$ without
changing the resolution of the lattice and thus without increasing the computation time and memory usage of our
simulations. Earlier \cite{Pogosian-Vachaspati:2000, Dvali-Valbuena:2022}, this erasure was only studied in the low-relativistic regime.

We developed a general ansatz with arbitrary monopole velocity $u_1$ and vacuum layer velocity $u_2$.
Lorentz boosting the vacuum layer solution yields
\begin{align*}
    \phi_\text{VL}(z)\rightarrow \tilde{\phi}_\text{VL}(z,t)=\phi_{\text{VL}}(\gamma_2 (z-u_2 t)).
\end{align*}
For the scalar field of the magnetic monopole solution, we have
\begin{align*}
    \phi_{\text{M}}(\vb*r)\rightarrow&\ \tilde{\phi}_\text{M}(\vb*r,t)=\phi_{\text{M}}(x,y,\gamma_1 (z-u_1 t)),
\end{align*}
where
$\gamma_1=\frac{1}{\sqrt {1-u_1^2}}$ and $\gamma_2=\frac{1}{\sqrt {1-u_2^2}}$ are the Lorentz factors for the magnetic
monopole and vacuum layer, respectively.
Since the gauge field is a Lorentz vector, it is necessary to apply the Lorentz transformation matrix to the vector
additionally to the transformation of the $z$-coordinate. This results in
\begin{align*}
    W^a_{\text{M}, \mu}(\vb*r)\rightarrow&\ \tilde{W}^a_{\text{M},\mu}(\vb*r, t)\\
    &=\begin{pmatrix}
                                                     -u_1 \gamma_1 W^a_{\text{M}, z}(x,y,\gamma_1 (z-u_1 t))\\
                                                     W^a_{\text{M}, x}(x,y,\gamma_1 (z-u_1 t))\\
                                                     W^a_{\text{M}, y}(x,y,\gamma_1 (z-u_1 t))\\
                                                     \gamma_1 W^a_{\text{M}, z}(x,y,\gamma_1 (z-u_1 t))
    \end{pmatrix}.
\end{align*}

For the combined initial configuration we use for the $\phi$ field, the product ansatz
\begin{align}
    \phi^a(\vb*r, t=0)&=\frac{1}{v}\tilde{\phi}^a_{\text{M}}(\vb*r,t=0)\tilde{\phi}_{\text{VL}}(z-d,t=0),\nonumber\\
    \partial_t \phi^a(\vb*r, t=0)&=\frac{1}{v}\partial_t \tilde{\phi}^a_{\text{M}}(\vb*r,t=0)\tilde{\phi}_{\text{VL}}(z-d,t=0)\nonumber\\
    &+\frac{1}{v}\tilde{\phi}^a_{\text{M}}(\vb*r, t=0)\partial_t \tilde{\phi}_{\text{VL}}(z-d,t=0),
\end{align}
where $d$ is the distance between the monopole and the vacuum layer.
For large enough distances, $d\gg m_h^{-1}$, the field $\phi^a$ goes to $\phi^a_\text{M}$ for $z<d/2$. 
For $z>d/2$, the field $\phi^a$ approaches the value 
$\phi_\text{VL} \hat{r}^a$. With our ansatz, there is no long-distance force between the monopole and the layer. We need to check the validity of this approximation.
In reality, for finite $d$ and $h$,  we expect several sources of interaction. Most significant is expected to be the quantum effect coming from the $SU(2)$ gauge bosons which acquire non-trivial mass profiles in the layer.
 
 First, let us assume that the $SU(2)$ theory stays in
 the perturbative weak coupling regime inside the layer. 
 The parameter regime justifying this assumption will be specified below.   
In such a case, the perturbative quantum effects will generate some $d$-dependent corrections to the magnetic field energy.
 
 This correction can be estimated as follows. 
 In the $U(1)$ invariant vacuum, the running gauge coupling 
 $g^2$ freezes 
 at the scale of the mass gap of the theory. This gap is given by the masses of gauge and Higgs bosons in this vacuum, $m_v \sim m_h$. 
 The effective low energy theory below this scale 
 is a theory of a free massless $U(1)$ Maxwell field.
 
 In the absence of the layer, the asymptotic value of the magnetic field energy density 
would be given by $|B^{U(1)}|^2 \rightarrow \frac{1}{g^2} \frac{1}{r^4}$. The presence of the $SU(2)$ invariant layer changes this energy 
in the following way.

Inside the $SU(2)$ invariant layer, the Higgs mass 
is essentially the same as in the $U(1)$ vacuum and is $\sim m_h$.  
The Higgs thereby decouples below this scale also in the effective theory inside the layer. 
However, not the gauge bosons. 
Since the Higgs VEV vanishes in the layer and we work in the regime $h  \gg m_v^{-1}$, 
the gauge coupling in the layer continues to evolve
all the way till the scale $h^{-1}$. This running is similar to the one in a
pure $SU(2)$ gauge theory. Since such a theory is asymptotically free, the evolved gauge coupling 
in the layer ($\equiv g_L^{2}$) is stronger than the gauge coupling in the exterior ($\equiv g_E^2$), $g_L^2 = g_E^2 + \delta g^2$. The difference is positive and is 
\begin{align}
\label{eq:deltag}
    \delta g^2=\frac{11}{12 \pi^2}g_E^4 \ln (m_v h)+\mathcal{O} (g_E^6).
\end{align} 

Thus, the presence of the layer decreases the magnetic energy of the monopole (see figure \ref{fig:magnetic_energy_density_quantum_effect}), resulting in an attractive potential between the two. Up to one-loop order, one can approximate it as
 \begin{equation} \label{LMpotential}
      V(d) \approx -\frac{11}{24 \pi}\frac{h}{d (d+h)}\ln (m_v h).
 \end{equation}\\
 
\noindent
\begin{minipage}{\linewidth}
  \includegraphics[width=\linewidth]{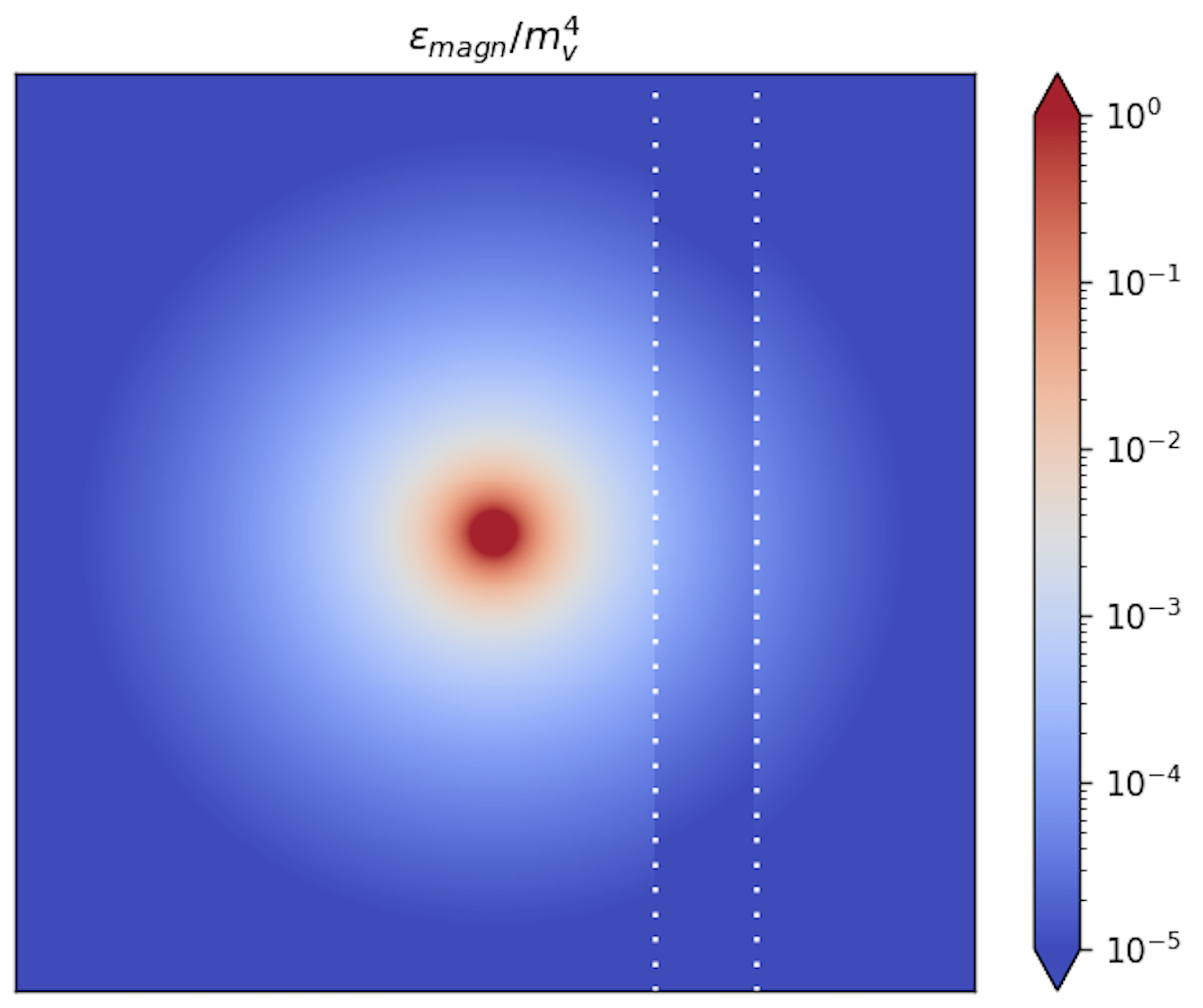}
  \captionof{figure}{The magnetic energy density of a magnetic monopole in the presence of an $SU(2)$ invariant vacuum layer taking into account the quantum correction \eqref{eq:deltag} for the coupling constant.}
  \label{fig:magnetic_energy_density_quantum_effect}
\end{minipage}\\[0.5cm]
 
 The acceleration of the monopole caused by the corresponding force is 
  $a \sim \frac{g_E^2 h}{m_v d^3} \ln(m_v h)$ for $h\ll d$.
  This force can be safely ignored at large distances. 
  Once the monopole enters the layer, the interaction is dominated by the classical profile of the Higgs field. This is explicitly taken into account by our numerical analysis.
 
 Let us now turn to the validity condition of the above-assumed perturbative weak 
 coupling regime inside the layer. This condition is rather simple. 
 Namely, the gauge coupling inside the layer must stop
 running before it hits the strong coupling scale of 
 the gauge $SU(2)$ theory, $\Lambda$.
 This gives us a condition,
 \begin{equation} \label{condconf}
 h^{-1} \gg \Lambda \,. 
\end{equation}

In the opposite case, $h^{-1} <  \Lambda$, the theory inside the layer will enter the strong coupling regime. The $SU(2)$ vacuum will become confining and generates a mass gap at the scale $\Lambda$. This leads to the effect of repelling the $U(1)$ electric flux
from the $SU(2)$ invariant vacuum towards the $U(1)$ invariant one, as originally studied in \cite{Dvali-Shifman:1996, Dvali-Vilenkin:2002, Dvali-Nielsen-Tetradis}.  Correspondingly, if the $SU(2)$ layer is thicker than the scale $\Lambda^{-1}$, the magnetic flux becomes screened in its interior. This effect is illustrated in figure \ref{fig:magnetic_and_electric_field_dual_superconductor}. 
We thereby work in a regime in which the thickness of the layer is much smaller than the scale of $SU(2)$ confinement.
Then, the quantum effects on the $U(1)$ field are reduced to the perturbatively-generated attractive potential (\ref{LMpotential}) between the monopole and the layer \cite{Dvali-Shifman:1996}.

Note that the layer will become a dual superconductor in the regime $h^{-1} <  \Lambda$. The magnetic field of the monopole will induce the 
surface charges that will screen the field inside the layer. 
However, the magnetic Gauss law will still hold.  The magnetic flux 
terminating on the surface charges from one side of the layer 
will be exactly equal to the flux originating from the opposite 
side. This regime goes beyond our numerical analysis and will not be considered.

Hence, we can use the following initial ansatz for the gauge fields.
\begin{align}
    W^a_{\mu}(\vb*r, t=0)&=\tilde{W}^a_{\text{M},\mu}(\vb*r, t=0),\\
    \partial_t W^a_{\mu}(\vb*r,t=0)&=\partial_t\tilde{W}^a_{\text{M},\mu}(\vb*r,t=0).
\end{align}
For the ansatz and the simulations, we take the Lorenz gauge $\partial_\mu W^\mu_a=0$.\\
\begin{minipage}{\linewidth}
  \includegraphics[width=\linewidth]{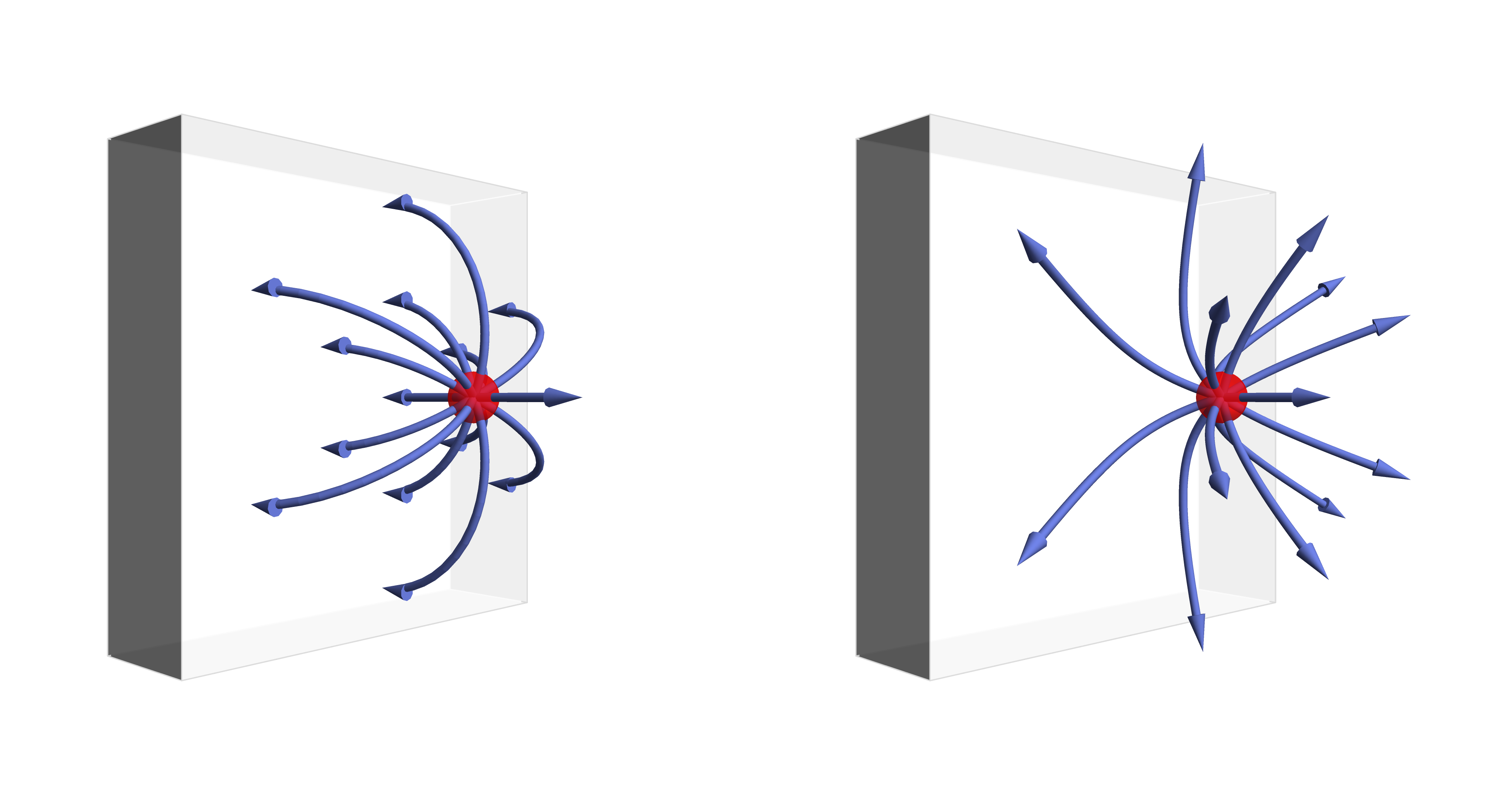}
  \captionof{figure}{In a superconductor layer, the electric field is screened. 
  Due to this, the electric flux lines terminate on the surface charges (left). At the same time, the magnetic flux lines are repelled (right). The SU(2) invariant vacuum represents a dual superconductor, and the behavior of magnetic and electric flux lines is reversed \cite{Dvali-Shifman:1996, Dvali-Nielsen-Tetradis}. Correspondingly, such a vacuum layer repels the electric flux while the magnetic flux terminates on surface magnetic image charges.}
\label{fig:magnetic_and_electric_field_dual_superconductor}
\end{minipage}\\[0.5cm]

Let us note that Ambj\o rn and Olesen pointed out in \cite{Ambjorn-Olesen:1988} that for an uniform  magnetic field $B>\frac{m_v^2}{g}$, the massive vector bosons 
can condense. 
This is happening because the magnetic field provides  
a bilinear term in the gauge fields that generates some 
imaginary frequency modes.
 This effect does not take place in the present case.
 
 Even though the Higgs profile vanishes inside the layer, 
the positive masses of the off-diagonal gauge bosons are still much higher than the negative contribution from the magnetic field. The latter, therefore, is insufficient for destabilizing 
the vacuum inside the layer.  
        \section{Numerical Implementation}\label{sec:numerical-implementation}
For the simulations, we used the programming language Python with the package Numba~\cite{Numba}, which translates our Python code into fast machine code.  Thereby this decreases the computation time substantially.

For a further increase of the computation speed and also an enhancement of the utilization of the working memory, we benefit from the axial symmetry of the system:
\begin{align}
    &\phi^1=x f_1 && \phi^2=y f_1 && \phi^3=z f_2\nonumber\\
    &W^1_x=xy f_3 && W^2_x=-x^2 f_3+f_4 && W^3_x=-y f_6\nonumber\\
    &W^1_y=y^2 f_3-f_4 && W^2_y=-xy f_3 && W^3_y=x f_6\nonumber\\
    &W^1_z=y f_5 && W^2_z=-x f_5 && W^3_z=0\nonumber\\
    & W^1_t=y f_7 && W^2_t=-x f_7 && W^3_t=0
\end{align}
where the functions $f_i$ depend only on the radius $r$ around the $z$-axis, $z$ and the time $t$.
With this method, it was sufficient to use only three lattice points in the $y$-direction.
The equations were solved on the $y=0$ plane, and for the neighboring planes, we used axial symmetry to determine the corresponding values of the fields.
This idea was adapted from an earlier paper by Pogosian and Vachaspati~\cite{Pogosian-Vachaspati:2000}.
The implementation of this symmetry was realized according to~\cite{Alcubierre:2001}.
The second iterative Crank-Nicolson method described in~\cite{Teukolsky:2000} was applied for the time evolution.

With the Python program, we analyze the following four cases:
\begin{align*}
    &(I)&& u_1 = 0 && u_2=-0.8\\
    &(II)&& u_1 = 0.4 && u_2 = 0\\
    &(III)&& u_1 = 0.8 && u_2 = 0\\
    &(IV)&& u_1 = 0.8 && u_2 = -0.98
\end{align*}
The first three cases will be used to study the electromagnetic radiation which gets emitted during the collision  between the monopole and the domain wall.
The fourth case serves as a simulation of the erasure mechanism for the ultra-relativistic regime with a
Lorentz factor of around $15$.

The lattice spacing in $x$- and $y$-direction was chosen to be $0.25m_v^{-1}$.
For the cases with monopole velocity $u_1 < 0.8$ the lattice spacing in $z$-direction was also $0.25m_v^{-1}$ and the
time step was set to $0.1m_v^{-1}$.
For the cases with monopole velocity $u_1 = 0.8$ we chose $0.125m_v^{-1}$ for the lattice spacing in $z$-direction and
$0.05m_v^{-1}$ for the time step.

For all four cases, we took the lattice size $[-60m_v^{-1}, 60m_v^{-1}]$ in the x-direction.
For $(I)$ the size in $z$-direction was chosen to be $[-60m_v^{-1}, 60m_v^{-1}]$ and for $(II)-(IV)$ we chose $[-30m_v^{-1}, 90m_v^{-1}]$. The time interval under investigation was $[0m_v^{-1}, 150m_v^{-1}]$.

The distance between the two domain walls of the vacuum layer was set to $h=20m_v^{-1}$, and the distance between the monopole and the vacuum layer was chosen to be $d=30m_v^{-1}$.
The constants $m_v$, $m_h$, and $g$ were set to one.
        \section{Results}\label{sec:results}
\begin{figure*}
  \centering
  \begin{subfigure}{0.30\textwidth}
    \includegraphics[trim=20 10 180 40,clip,width=\textwidth]{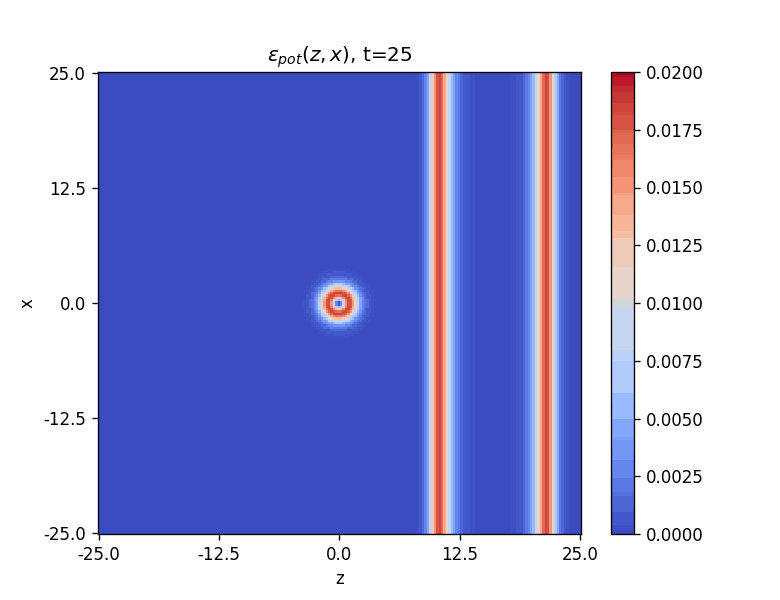}
  \end{subfigure}
  \begin{subfigure}{0.29\textwidth}
    \includegraphics[trim=35 10 180 40,clip,width=\textwidth]{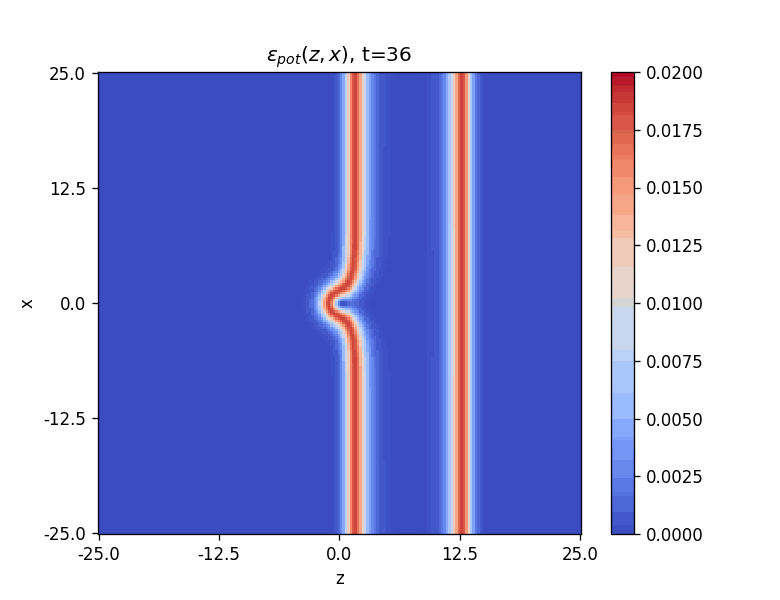}
  \end{subfigure}
  \begin{subfigure}{0.36\textwidth}
    \includegraphics[trim=35 10 40 40,clip,width=\textwidth]{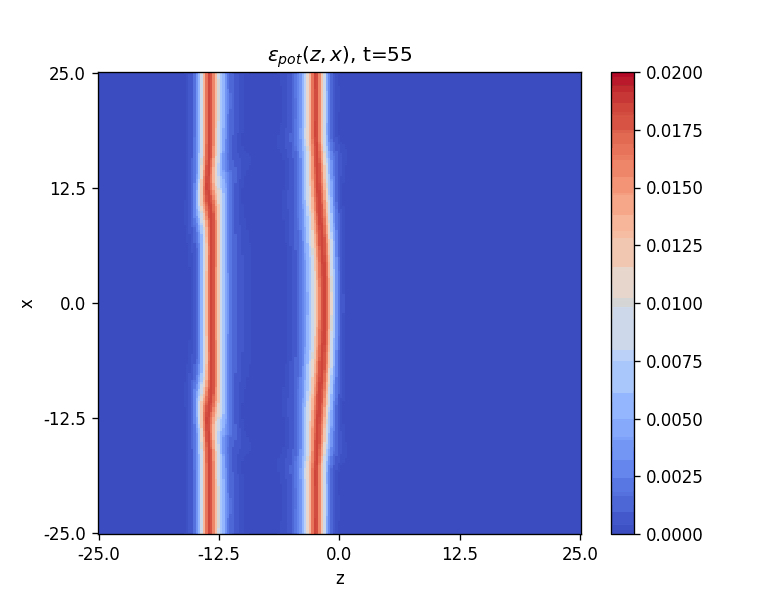}
  \end{subfigure}
  \caption{Evolution of the potential energy density for the case $(I)$ in the $y=0$ plane.
  The length and time values are in units
  of $m_v^{-1}$, and the energy density in units of $\frac{m_v^4}{g^2}$. 
  The vacuum layer moves over the monopole and unwinds it. Furthermore, we can observe radial disturbances that move along the first domain wall with the speed of light. The second domain wall also shows some deformations through the backreaction of the emitted radiation.}
  \label{fig:animation_potential_energy_density_case_I}
\end{figure*}

\begin{figure*}
  \centering
  \begin{subfigure}{0.31\textwidth}
    \includegraphics[trim=20 10 180 40,clip,width=\textwidth]{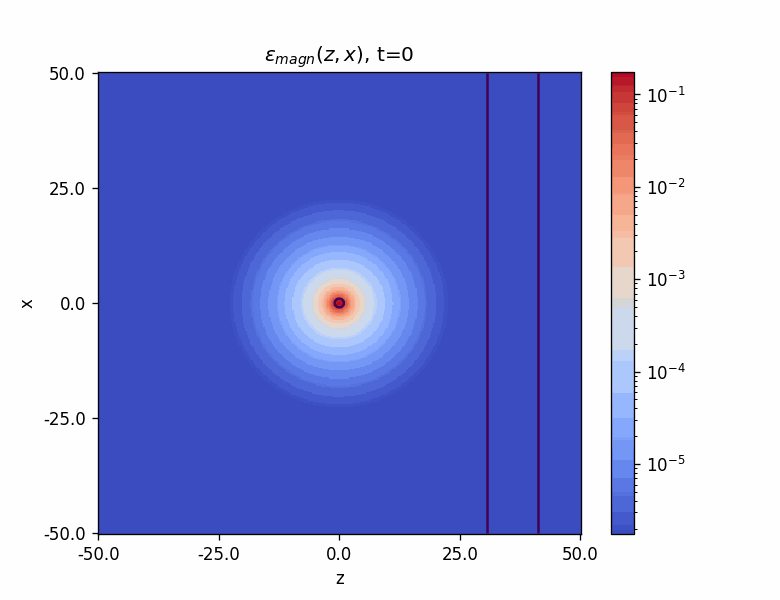}
  \end{subfigure}
  \begin{subfigure}{0.30\textwidth}
    \includegraphics[trim=35 10 180 40,clip,width=\textwidth]{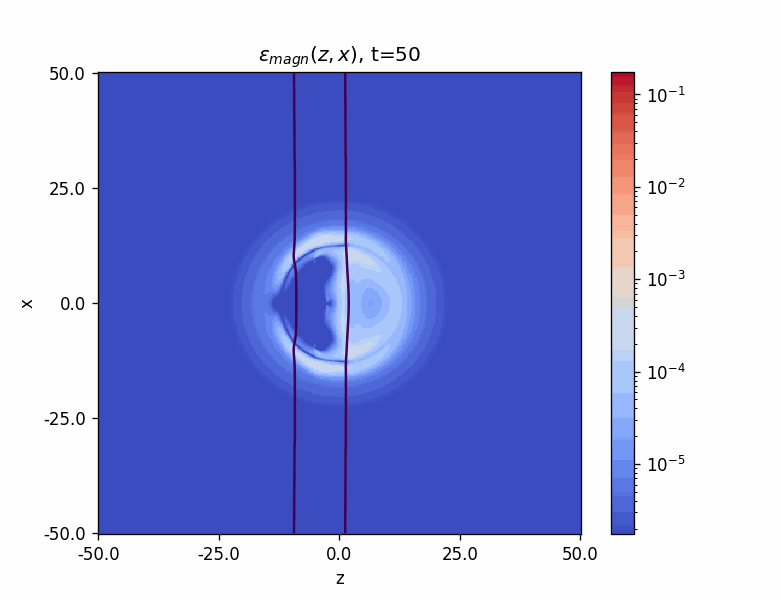}
  \end{subfigure}
  \begin{subfigure}{0.37\textwidth}
    \includegraphics[trim=35 10 40 40,clip,width=\textwidth]{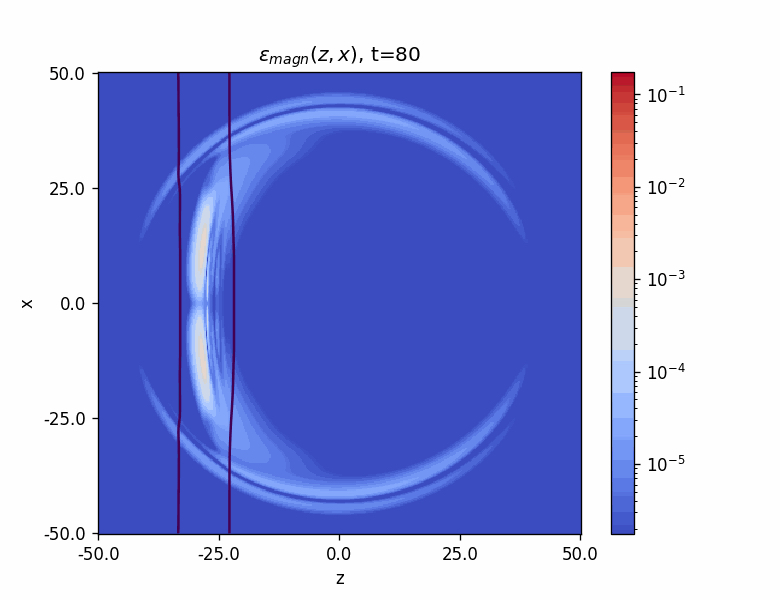}
  \end{subfigure}
  \caption{Evolution of the magnetic energy density for the case $(I)$ in the $y=0$ plane.
  The length and time values are in units
  of $m_v^{-1}$, and the energy density is in units of $\frac{m_v^4}{g^2}$.
  The black lines illustrate the $SU(2)$ invariant vacuum layer. We used the value $\sqrt{\phi^a \phi^a}=0.5$ to draw its contour.
  As we can see,  after the collision between the vacuum layer and the magnetic monopole, part of the magnetic energy moves away radially. In contrast, most of the magnetic energy is captured within the layer where the magnetic charge spreads.}

\label{fig:animation_magnetic_energy_density_case_I}
\end{figure*}

\begin{figure*}
  \centering
  \begin{subfigure}{0.31\textwidth}
    \includegraphics[trim=20 10 180 40,clip,width=\textwidth]{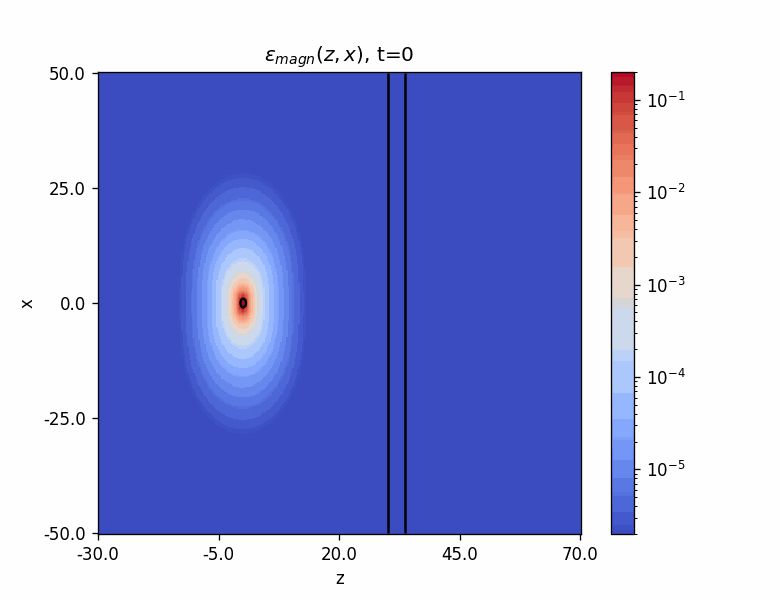}
  \end{subfigure}
  \begin{subfigure}{0.30\textwidth}
    \includegraphics[trim=35 10 180 40,clip,width=\textwidth]{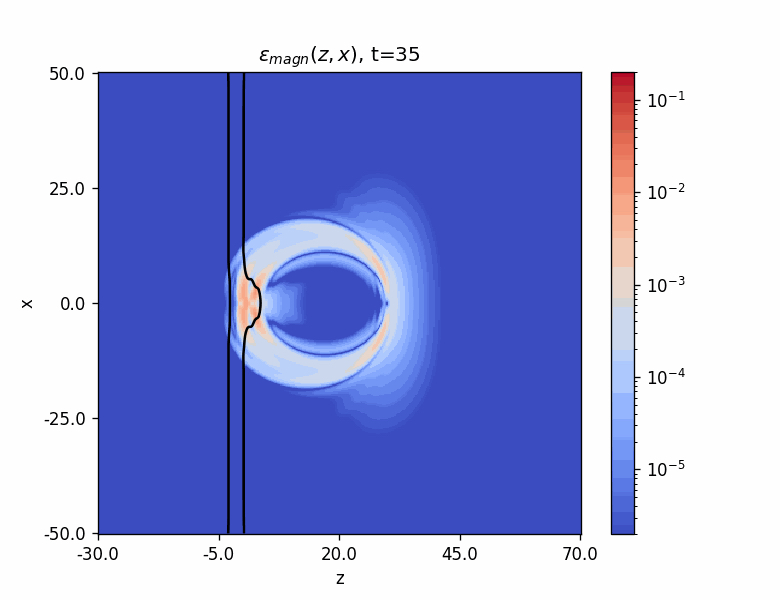}
  \end{subfigure}
  \begin{subfigure}{0.37\textwidth}
    \includegraphics[trim=35 10 40 40,clip,width=\textwidth]{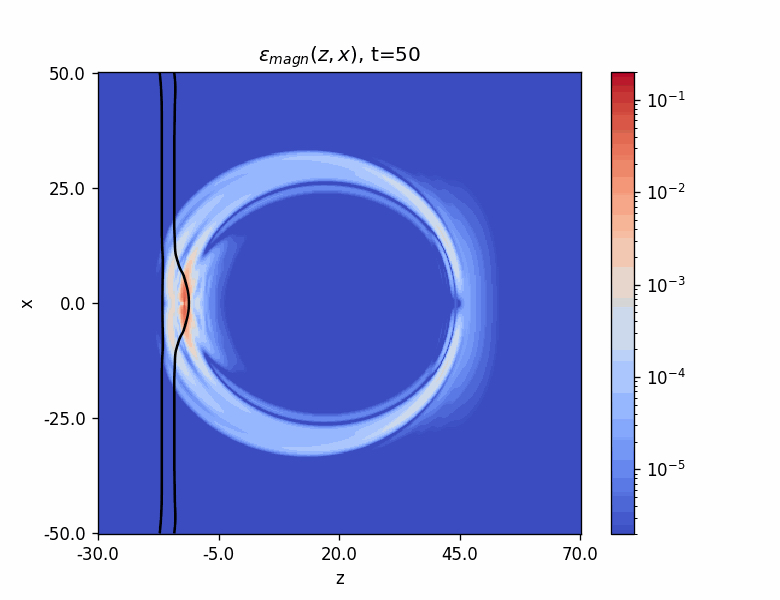}
  \end{subfigure}
  \caption{Evolution of the magnetic energy density for the case $(IV)$ in the $y=0$ plane.
  The length and time values are in units of $m_v^{-1}$, and the energy density is in units of $\frac{m_v^4}{g^2}$.
  Again, the black lines illustrate the $SU(2)$ invariant vacuum layer. We observe the same behavior as for case $(I)$.
  The magnetic energy of the monopole unwinds, the remaining energy moves away radially, and most of the energy is captured within the two domain walls. One further particular detail can be extracted from these figures. The magnetic energy is not erased immediately everywhere. It takes a finite time for the magnetic field to respond to the spread of the magnetic source. An electromagnetic pulse transports the information about the erasure.}
  \label{fig:animation_magnetic_energy_density_case_IV}
\end{figure*}

In all four cases, $(I)-(IV)$, we observe the erasure of the magnetic monopole during the collision with the vacuum layer.
For the case $(I)$, some frames of the evolution of the potential energy density and magnetic energy density can be found in figures~\ref{fig:animation_potential_energy_density_case_I} and~\ref{fig:animation_magnetic_energy_density_case_I},
respectively.
For the ultra-relativistic case $(IV)$, the evolution of the magnetic energy density is plotted in figure ~\ref{fig:animation_magnetic_energy_density_case_IV}.
With this, we checked the DLV mechanism~\cite{Dvali-Liu-Vachaspati:1997} for the $SU(2)$
gauge theory with $\phi^6$ potential for low relativistic and ultra-relativistic collision velocities.
Additionally to the figures, the results of the numerical simulations can be found in the following video:\\
\url{https://youtu.be/JZaXUYikQbo}
\begin{figure*}
  \centering
  \begin{subfigure}{0.30\textwidth}
    \includegraphics[trim=40 20 360 40,clip,width=\textwidth]{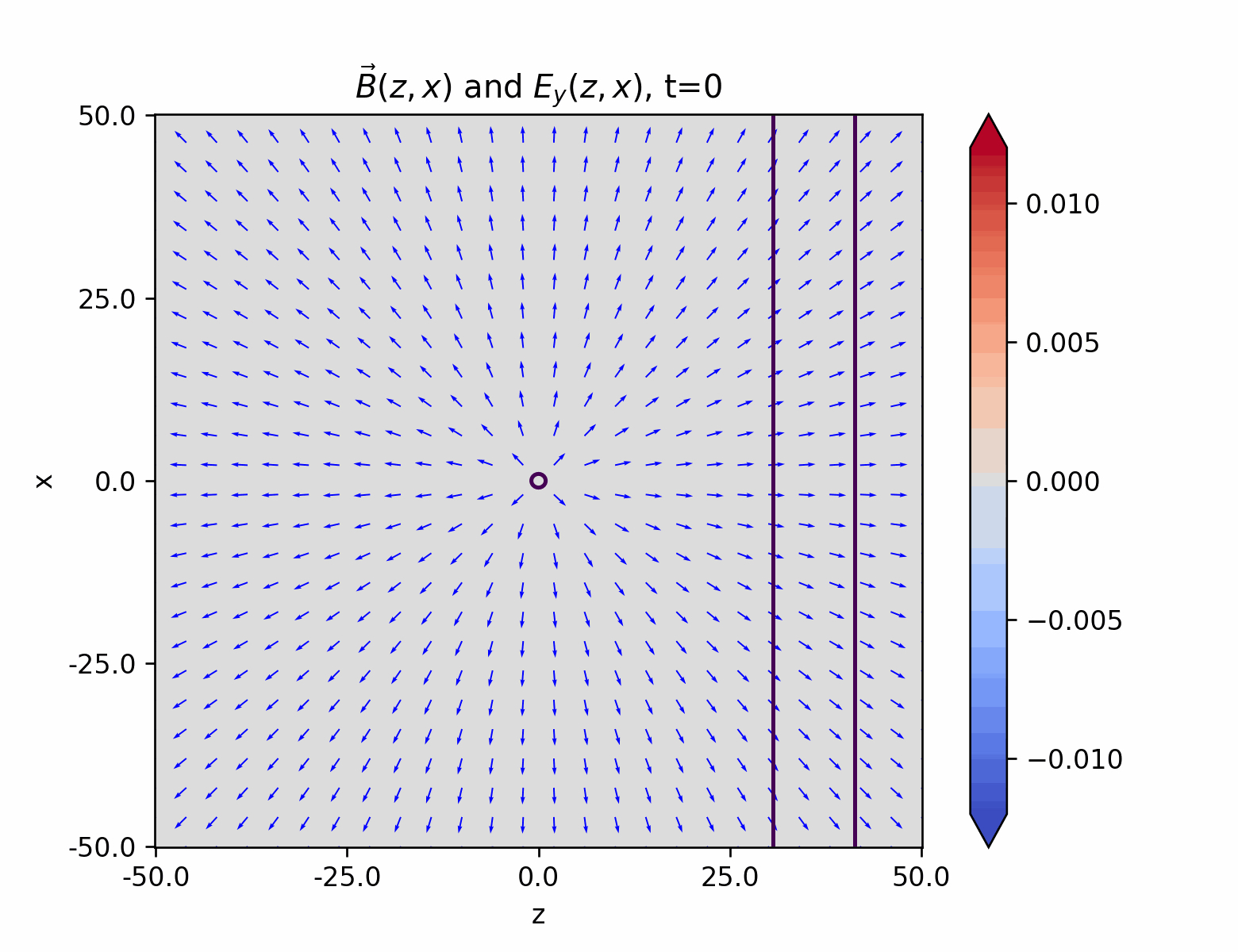}
  \end{subfigure}
  \begin{subfigure}{0.29\textwidth}
    \includegraphics[trim=67 20 360 40,clip,width=\textwidth]{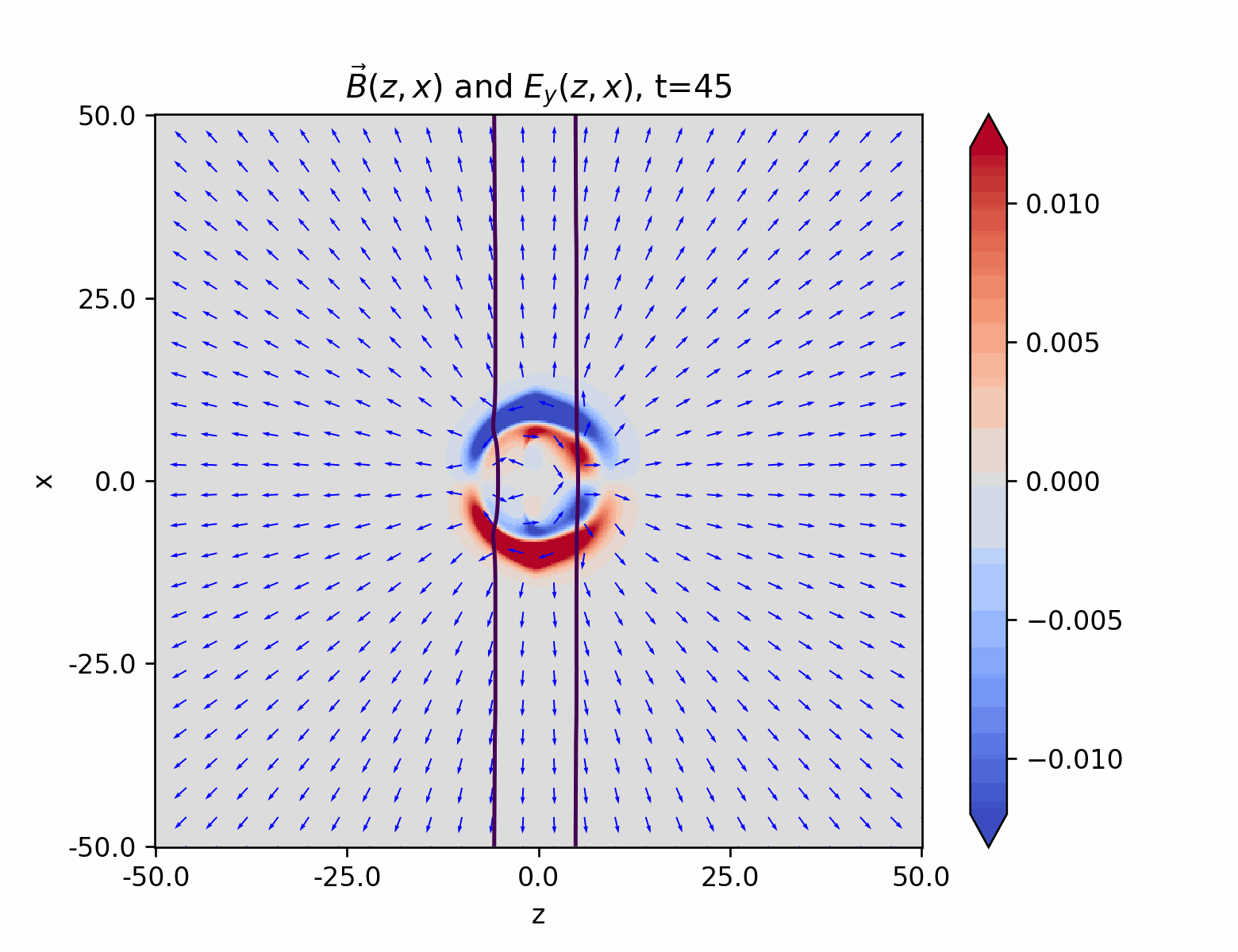}
  \end{subfigure}
  \begin{subfigure}{0.35\textwidth}
    \includegraphics[trim=67 20 120 40,clip,width=\textwidth]{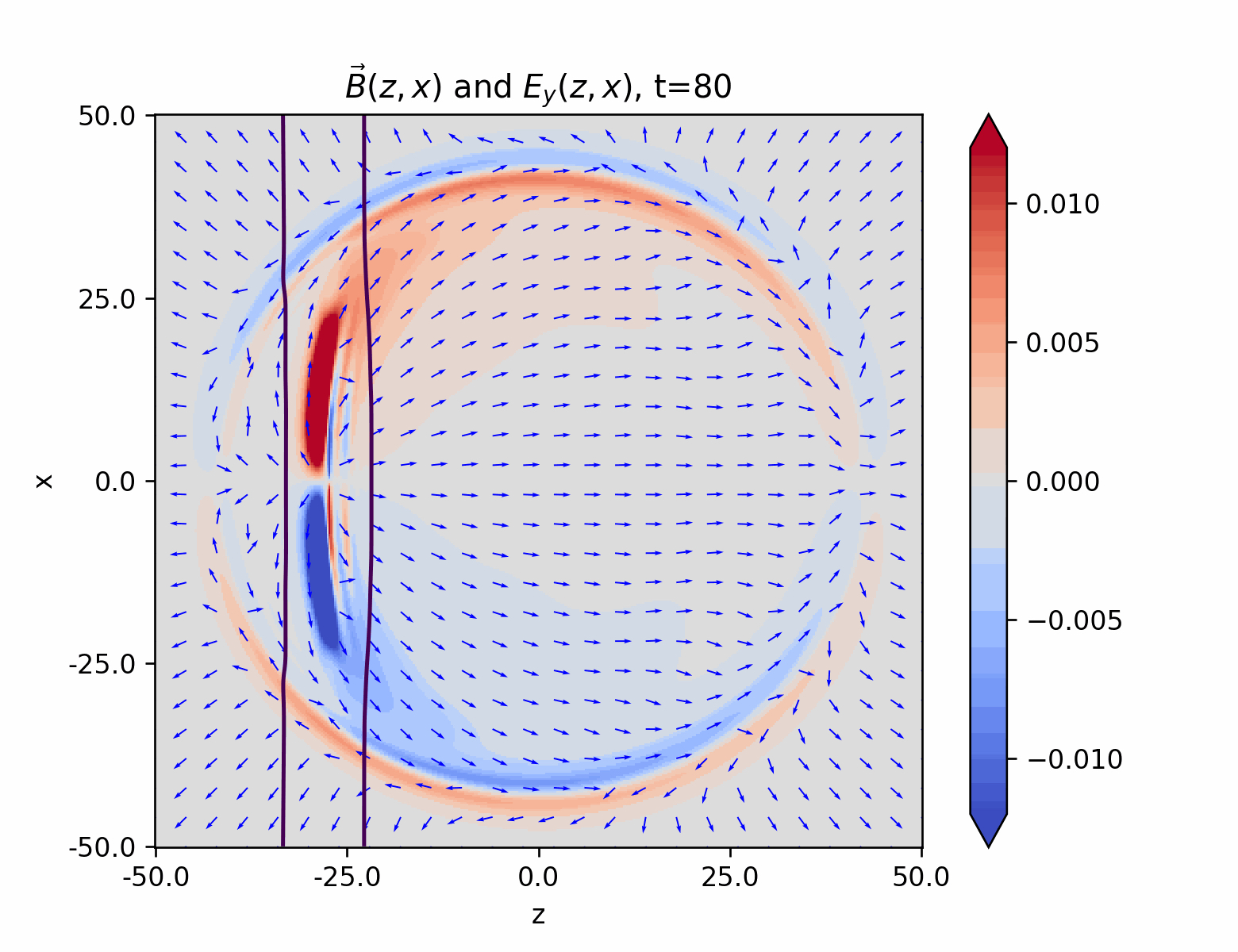}
  \end{subfigure}
  \caption{Evolution of the magnetic and electric field for the case $(I)$ in the $y=0$ plane.
  The length and time values are in units
  of $m_v^{-1}$. The arrows illustrate the direction of the magnetic field, whereas the red and blue colors illustrate the electric field. Red colors represent positive values of $E_y$, and blue colors
  represent negative values of $E_y$. Initially, the magnetic field arrows point radially away from the origin, where the magnetic monopole is located. After the collision, the arrows adjust to the positive $z$-direction when the  electromagnetic pulse moves over them. From the axial symmetry of the system, we can conclude that the electric field lines are circles around the $z$-axis that extend with time.}
  \label{fig:animation_magnetic_electric_field_case_I}
\end{figure*}

  Note that in figures \ref{fig:animation_magnetic_energy_density_case_I}, \ref{fig:animation_magnetic_energy_density_case_IV} there is spherical energy radiation with a factor of around $10^{-3}$ smaller than the energy density in the magnetic monopole's core. 
This observation is valid for all the considered monopole and vacuum layer velocities.
This radiation spreads at the speed of light and corresponds to electromagnetic radiation. We confirmed it by analyzing the Fourier spectrum of the pulse.

Before we continue with the investigation of the form of electromagnetic radiation, we give some more comments on the phenomena of erasure itself. As we mentioned, the magnetic monopole is always erased, and there is no evidence to suggest that it could pass through the vacuum layer, even in the ultra-relativistic regime. This phenomenon can be attributed to the loss of coherence \cite{Dvali-Liu-Vachaspati:1997}. After the collision, most of the coherence is carried away by the radiation. This line of reasoning has already been presented in previous studies about monopole anti-monopole annihilation \cite{Dvali-Valbuena-Zantedeschi:2022} and vortex erasure \cite{Dvali-Valbuena:2022}.

Furthermore, this behavior is also explained by entropy arguments.
A state with radiation has more entropy than a state with a monopole.
The entropy of a monopole is significantly lower than the entropy needed to saturate the unitarity bound~\cite{Dvali:2020}, and thus the recreation of a monopole is strongly suppressed.

To characterize the identified electromagnetic radiation, we can study the direction of its magnetic and electric fields.
In figure~\ref{fig:animation_magnetic_electric_field_case_I}, some frames of the evolution of the magnetic and electric field are depicted.
Before the collision, the magnetic field pointed radially away from the center where the monopole was located.
After the layer passes over the monopole, the magnetic field shifts in the direction toward the positive side of the $z$-axis. This shift proceeds at the speed of light and is a consequence of the appearance of an induced current during the interaction process.
The current flows in circles around the $z$-axis, leading to a magnetic field perpendicular to the wall, i.e.~parallel to the $z$-axis.
During the erasure of the monopole, an electric field emerges and spreads away radially.
In the $y=0$ plane, the electric field points only in the $y$-direction.
From the axial symmetry of our system, we can conclude that the electric field lines are circles around the $z$-axis.
The outer electric field of the pulse points anti-clockwise around the $z$-axis, whereas the inner electric field points clockwise around the $z$-axis.

The magnetic field arrows (see figure \ref{fig:magnetic_electric_field_case_I_time75}) wriggle in a banana-shaped form around the pulse's center.
Although the interaction analyzed here is a combined process of the erasure and acceleration of a magnetic mo-

\noindent
\begin{minipage}{\linewidth}
  \includegraphics[trim=40 0 80 0,clip,width=\linewidth]{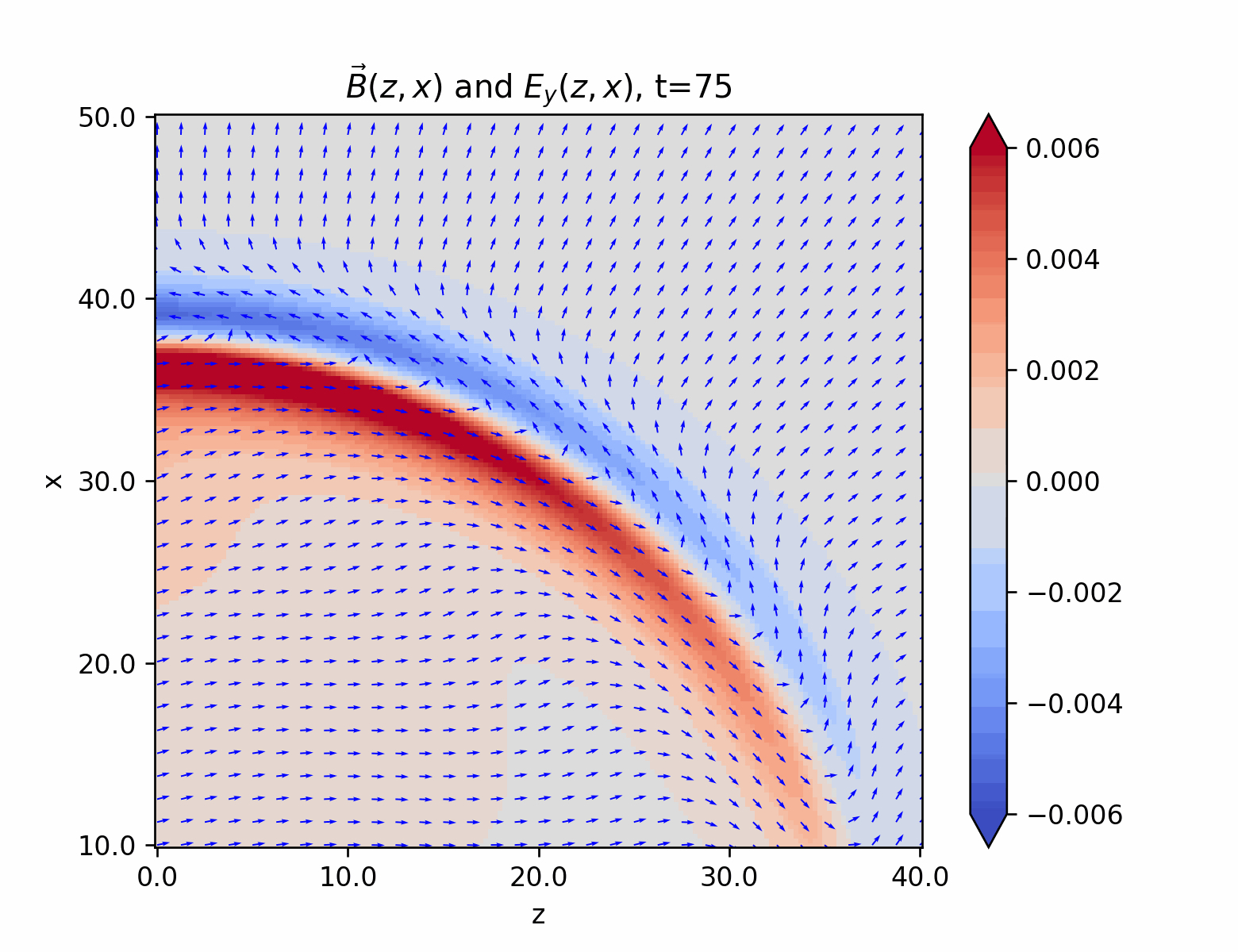}
  \captionof{figure}{Magnetic and electric field for case $(I)$ at time $t=75m_v^{-1}$. The length and time values are given in $m_v^{-1}$ units.}
  \label{fig:magnetic_electric_field_case_I_time75}
\end{minipage}\\[0.5cm]

\noindent
nopole, and the magnetic charge is not located at one point, the behavior of electromagnetic radiation is qualitatively the same as for an accelerated magnetic point charge.

The previous observations prompted us to reconstruct a radiation pattern for different initial monopole velocities to compare it with equation \eqref{eq:energy-radiation-pattern}.
We approximated the center of radiation emission using the radiation energy density data. 
Furthermore, we integrated the radiation energy density over the pulse and created a radiation pattern to see in which direction most radiation gets emitted.

For the cases $(I)$, $(II)$ and $(III)$, we chose the times $85m_v^{-1}$, $110m_v^{-1}$ and $85m_v^{-1}$ respectively and created out of the data for the electromagnetic energy density $\varepsilon=\frac{1}{2}\abs{\vb*E^\text{U(1)}}^2+\frac{1}{2}\abs{\vb*B^\text{U(1)}}^2$
the radiation patterns at these moments of time.
The results are given in figure~\ref{fig:radiation_pattern_integrated}.

The loops are not bent in the same way as in the case of an accelerated point charge, given
in figure~\ref{fig:radiation_pattern_theory}.
Nevertheless, qualitatively the behavior of the angle $\theta_{\max}$ corresponding to the maximum of radiation emission is conformable to equation \eqref{eq:angle_maximal_radiation}, describing the radiation emitted by an accelerated point charge.
This behavior is independent of the velocity of the vacuum layer.

        \section{Conclusion and Outlook}\label{sec:conclusion-and-outlook}
In this work, we bear out the
DLV mechanism of erasure of magnetic monopoles by domain walls~\cite{Dvali-Liu-Vachaspati:1997}. 
 We performed our numerical study on a prototype model 
 with $SU(2)$ gauge symmetry, which possesses the degenerate vacua with $U(1)$ and $SU(2)$ invariant phases
 ~\cite{Dvali-Vilenkin:2002, Dvali-Nielsen-Tetradis}.\\
  \begin{minipage}{\linewidth}
  \includegraphics[width=\linewidth]{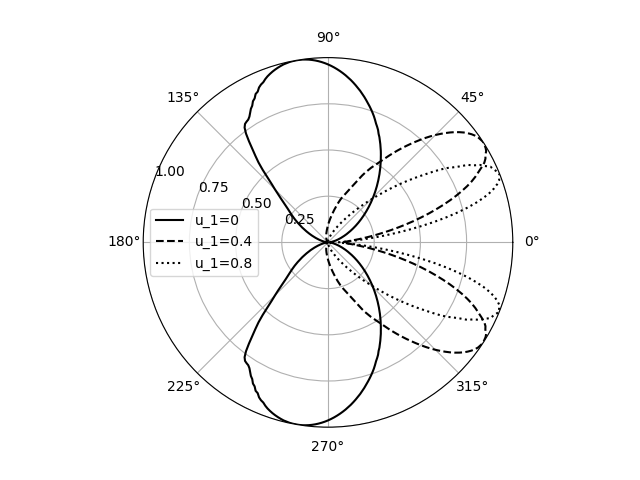}
  \captionof{figure}{Radiation patterns for the cases $(I)$, $(II)$ and $(III)$ at some moments after the collision
  between the magnetic monopole and the $SU(2)$ invariant vacuum layer.
  The radius represents the normalized value of the radiation energy $\frac{E}{E_{\max}}$.}
  \label{fig:radiation_pattern_integrated}
\end{minipage}\\[0.5cm]

   Correspondingly, it has a solution in the form of the layer of an $SU(2)$ invariant vacuum, 'sandwiched' in between the $U(1)$ invariant vacua.  The layer is taken to be sufficiently thin so that the effects of the
 $SU(2)$ confinement on the gauge fields, discussed 
 in~\cite{Dvali-Shifman:1996, Dvali-Nielsen-Tetradis}, can be ignored.
 
  The $U(1)$ vacua support the 
 't Hooft-Polyakov magnetic monopoles.  When a monopole meets the 
 wall, it gets erased, and the magnetic charge spreads 
 in the layer.  
   We study the process of the erasure numerically. 
   Special attention is paid to the emission of electromagnetic radiation.  Remarkably, our simulations 
 allow us to analyze the radiation 
   dynamics convincingly, despite its relatively low energy.  
The radiation emission resembles the radiation emitted due to the acceleration of a magnetic point charge.
We noted these similarities in the shape of the electric and magnetic fields and the form of the radiation pattern.

This paper serves as proof of principle and motivation for future work, as it is a way to characterize and extract possible observables of the DLV mechanism.

Given that this mechanism is an occurrence in the early universe, it could have relevant effects on the cosmic microwave background. Studies in this direction already exist for cosmic strings \cite{Vilenkin:1984} and domain wall networks \cite{Lazanu:2015}. Additionally, the erasure of defects may contribute to the emission of high-energy particles in the early universe, similar to the study of radiation in monopoles and anti-monopoles connected by strings~\cite{Berezinsky:1997kd}.

Furthermore,
our analysis of the erasure mechanics can be 
straightforwardly generalized to larger symmetry groups. 

The next step is to consider the study of gravitational radiation from the erasure of topological defects. It is a new mechanism that 
gives relevant imprints to the known scenarios of gravitational wave emission from phase transitions in the early universe (for a review see for instance \cite{Caprini:2019egz}).
In this direction, the gravitational radiation from topological defects was  previously studied in the context of monopoles connected by strings. Originally, this was  
performed by Martin and Vilenkin in point-like approximation~\cite{Martin:1996}. A more recent study~\cite{Dvali-Valbuena-Zantedeschi:2022}, which goes beyond this approximation, reveals that in the regime of comparable widths of strings and monopoles, the monopole and anti-monopole never go through one another and oscillate.  Instead, they get directly erased (annihilated) in a single collision, converting the entire energy into the waves of Higgs, gauge, and gravitational fields. 
In the present analysis of wall-monopole collision, a similar maximal rate of erasure is observed.  
Due to this, we expect a high efficiency of gravitational wave production during the erasure. This will be studied elsewhere.
        \section*{Acknowledgements}

This work was supported in part by the Humboldt Foundation under Humboldt Professorship Award, 
by the European Research Council Gravities Horizon Grant AO number: 850 173-6,  by the Deutsche Forschungsgemeinschaft (DFG, German Research Foundation) under Germany's Excellence Strategy - EXC-2111 - 390814868, and Germany's Excellence Strategy under Excellence Cluster Origins.\\

\noindent
\textbf{Disclaimer:}
Funded by the European Union. Views and opinions expressed are however those of the authors only and do not necessarily reflect those of the European Union or European Research Council. Neither the European Union nor the granting authority can be held responsible for them.

        \renewcommand*{\bibfont}{\footnotesize}
        \printbibliography
    \end{multicols}

\end{document}